Scientific Research Publishing

# How Good Is the Debye Model for Nanocrystals?

## Enrique N. Miranda[1,2*], Geraudys Mora-Barzaga[1]

[1]CONICET and Facultad de Ciencias Exactas y Naturales, Universidad Nacional de Cuyo, Mendoza, Argentina
[2]IANIGLA, CONICET, Mendoza, Argentina
Email: *emiranda@mendoza-conicet.gov.ar

## Abstract

The question here is whether the Debye model is suited to evaluate the specific heat of nanocrystals. For this, the simplest possible nanocrystal is considered: a basic cubic structure made of atoms that interact through a harmonic potential. This simple model can be solved exactly. This allows the dispersion relation of the mechanical waves to be determined, so that calculating the exact specific heat turns out to be quite straightforward. Then, the same problem is solved using the Debye approximation. Our findings show that the specific heat of a nanocrystal evaluated exactly is higher than the value found in the thermodynamic limit, that is to say, the specific heat decreases as the nanocrystal size increases. In addition, it becomes clear that the Debye model is a poor approximation for calculating the specific heat of a nanocrystal. Naturally, the Einstein model yields an even worse result. The cause of the discrepancy is found in the role of the nanocrystal surface.

## 1. Introduction

The boom of physics at a nanoscopic scale justifies questioning the validity of some of the usual claims when working with systems made up of a few hundred or a few thousand particles. Previous papers [1] [2] dealt with studies on how the thermodynamic properties change when evaluated following the different ensembles of statistical mechanics. These demonstrated that, for systems with very few thousand particles, the different thermodynamic magnitudes reach the same values independently of the ensemble considered. In this sense, a claim of conventional wisdom was confirmed. However, for systems with tens of particles, it was also shown that some fundamental statements ceased to be valid; for example, the specific heat depends on the context in which it is measured [2].

In addition, the question arises whether the usual models for solids are applicable at the nanoscale. A recent article [3] introduced "The Einstein nanocrystal", which is the simplest model for a nanoparticle. The atoms are assumed to be configured in a simple cubic lattice, where each can oscillate in three different directions and with the same frequency in all cases. Such a straightforward model served to reveal the importance of the surface for the specific heat of the nanocrystal. The present article takes the next step and introduces a similar nanocrystal consisting of a simple cubic lattice where the atoms interact through a harmonic potential ("springs"), and there are normal modes with certain distribution of frequencies. The objective here is to test the validity of the approximations made using the very well-known Debye solid [4] [5] to study systems with a few hundred or a few thousand atoms.

The body of this article is structured in three main sections. Section 2 presents the model of a nanocrystal formed by $N_A$ atoms and shows its exact solution. Section 3 introduces the approximations made to obtain the specific heat of the Debye solid and, for comparison purposes, the Einstein model is also included. By comparing both models with the exact results, it will become clear that they are not suited for systems with thousands of atoms. Finally, Section 4 summarizes the conclusions.

## 2. The Simplest Nanocrystal Model

**Figure 1** shows the proposed nanocrystal model. It is a simple cubic structure with a lattice constant of $a_0$. There are a total of $N_A$ atoms, which means that the side of the nanocrystal has $n_L = N_A^{1/3}$ atoms and the total volume is given by $V = (n_L - 1)^3 a_0^3$. The atoms with mass $M$ are connected to each other through springs characterized by a constant $K$.

It should be noticed that we deal with a toy model since there are no simple cubic crystals in nature. We use this crystal structure because our concern is to check the validity of the Debye model at the nanoscale and do not pretend to reproduce any actual experimental result.

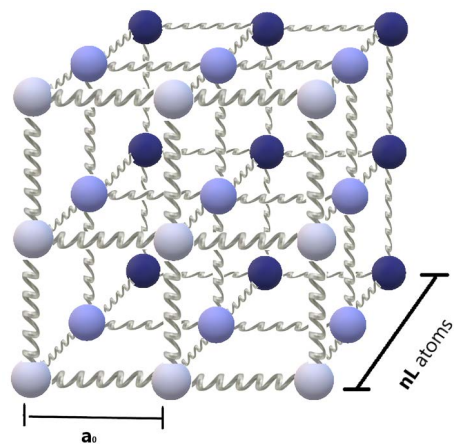

**Figure 1.** The nanocrystal model studied. A simple cubic structure with $n_L$ atoms per side and a lattice constant of $a_0$.

In terms of classical mechanics, the behavior of this system is well-known: the system has $3n_L^3$ different oscillation modes whose frequencies can be calculated easily. The case of a one-dimensional lattice with pinned boundary conditions is dealt with in textbooks [5]. Each oscillation mode is characterized by a wave vector $\boldsymbol{k}$, whose module is $k = m\pi/(n_L a_0)$, where $m = 0, 1, 2, \cdots, n_L - 1$ and the frequency $\omega$ is given by:

$$\omega^2(\boldsymbol{k}) = \left(\frac{4K}{M}\right)\sin^2\left(\frac{ka_0}{2}\right) \tag{1}$$

Generalizing to the three-dimensional case [6] leads to the following relation between the wave vector $\boldsymbol{k}$ and the frequency $\omega$:

$$\boldsymbol{k} = \frac{\pi}{n_L a_0}\left(m_x, m_y, m_z\right) \tag{2}$$

$$\omega^2 = \frac{4K}{M}\left(\sin^2\left[\frac{\pi}{2n_L}m_x\right] + \sin^2\left[\frac{\pi}{2n_L}m_y\right] + \sin^2\left[\frac{\pi}{2n_L}m_z\right]\right) \tag{3}$$

where $m_x, m_y, m_z = 0, 1, 2, \cdots, n_L - 1$. The characteristic frequency of the system is:

$$\omega_0 = 2\sqrt{K/M} \ . \tag{4}$$

It is necessary to consider all possible combinations of $m_x$, $m_y$ and $m_z$, except (0, 0, 0) that obviously corresponds to the body at rest.

So far for the classical treatment, in quantum terms, these results are interpreted as the system having phonons that are characterized by a wave vector $\boldsymbol{k}$ and a frequency $\omega(\boldsymbol{k})$, and that obey Bose-Einstein's statistics. It is a usual exercise in any statistical mechanics course [5] to show that the energy of such a system is given by:

$$U = \sum_k \left(\frac{1}{2}\hbar\omega(\boldsymbol{k}) + \frac{\hbar\omega(\boldsymbol{k})}{e^{\frac{\hbar\omega(\boldsymbol{k})}{k_B T}} - 1}\right) \tag{5}$$

It follows immediately that the heat capacity $C = \mathrm{d}U/\mathrm{d}T$ is:

$$C = \sum_k \left(\frac{\hbar^2\omega^2(\boldsymbol{k})e^{\frac{\hbar\omega(\boldsymbol{k})}{k_B T}}}{k_B T^2 \left(e^{\frac{\hbar\omega(\boldsymbol{k})}{k_B T}} - 1\right)^2}\right) \tag{6}$$

The sum of all the $\boldsymbol{k}$ vectors can be broken down into a triple sum: $\sum_{\boldsymbol{k}}(\ ) \rightarrow \sum_{k_x}\sum_{k_y}\sum_{k_z}(\ )$, and one should remember that there are three possible polarization states (two transverse and one longitudinal) of each phonon.

Also, some abbreviations are included to simplify the notation. We make use of an auxiliary function defined as follows:

$$R\left(m_x, m_y, m_z\right) = \left(\sin^2\left[\frac{\pi}{2n_L}m_x\right] + \sin^2\left[\frac{\pi}{2n_L}m_y\right] + \sin^2\left[\frac{\pi}{2n_L}m_z\right]\right)^{1/2} \tag{7}$$

The characteristic frequency of the system is $\omega_0 = 2(K/M)^{1/2}$. The temperature can be expressed in terms of a dimensionless variable $t$ as $T = t T_c$ with $T_c = \hbar \omega_0 / k_B$. Keeping in mind all these definitions and remembering that there are $n_L^3$ atoms, the specific heat per atom $c$ is written:

$$c_{ex} = \frac{3 k_B}{n_L^3} \sum_{m_x=0}^{n_L-1} \sum_{m_y=0}^{n_L-1} \sum_{m_z=0}^{n_L-1} \frac{R^2(m_x, m_y, m_z) e^{R/t}}{t^2 (e^{R/t} - 1)^2}; \quad m_x + m_y + m_z \neq 0 \quad (8)$$

In the rest of this article, the specific heat is expressed in terms of units of $k_B$. In this way, $c$ becomes an adimensional variable as well as the temperature $t$.

This is expression (8) is exact, noted by subscript "*ex*", and it gives the specific heat per atom for a nanocrystal with a simple cubic structure and $n_L$ atoms per side. The first question that arises is how this parameter changes according to the size of the system. And the answer is that it changes very slowly, as seen in **Figure 2**.

In the next section, the specific heat will be calculated again but introducing step-by-step the approximations of the Debye model.

## 3. The Debye Approximation

Actually, several approximations are made in the Debye model for solids [4] [5]. First, the relation between the frequency and the wave vector is linearized. Then, a frequency continuum is assumed and a maximum cutoff frequency is introduced.

Linearizing the dispersion relation of the phonons implies approximating the sine that appears in (1) by its argument. In one dimension:

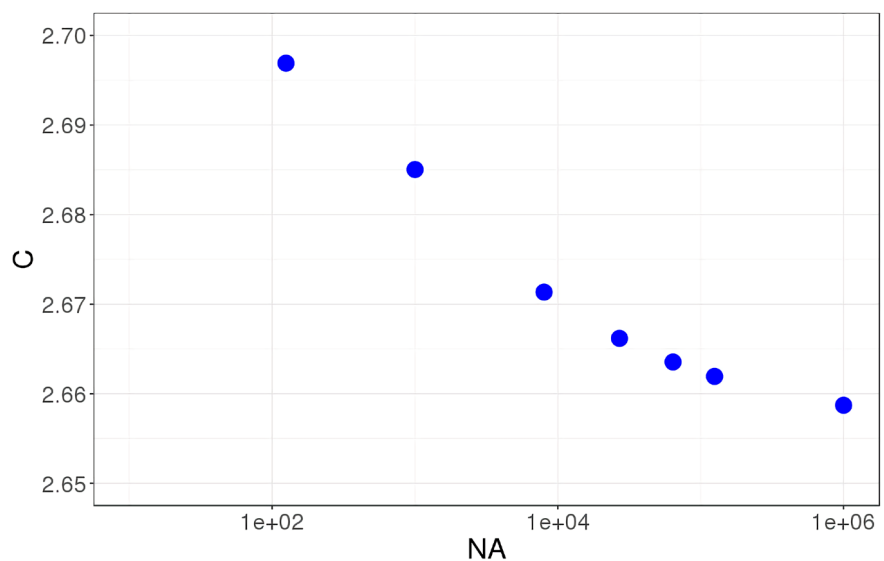

**Figure 2.** Specific heat per atom $c$ evaluated exactly for $t = 1$ in relation to the number of atoms $N_A$. It is clear that the specific heat for a small nanocrystal is higher than the value in the thermodynamic limit and that it converges slowly towards that value. The specific heat is in units of $k_B$.

$$\omega(k) = 2\left(\frac{K}{m}\right)^{1/2}\left|\sin\left(\frac{ka_0}{2}\right)\right| \cong \omega_0 \frac{ka_0}{2} = \omega_0 \frac{\pi m}{2n_L} \qquad (9)$$

with $m = 0, 1, 2, \cdots, n_L - 1$.

In three dimensions, the auxiliary function $R(m_x, m_y, m_z)$ defined previously needs to be modified as follows:

$$R'(m_x, m_y, m_z) = \left(\left(\frac{\pi}{2n_L}m_x\right)^2 + \left(\frac{\pi}{2n_L}m_y\right)^2 + \left(\frac{\pi}{2n_L}m_z\right)^2\right)^{\frac{1}{2}} \qquad (10)$$

To calculate the specific heat, the same expression (8) is used, but with function $R'$ given by (10).

Then, the frequencies are assumed to be continuous and the sum in (6) is replaced by an integral that includes the density of states for phonons. And finally, a maximum frequency, known as the Debye frequency, is introduced. Physically, the existence of a maximum frequency is determined by the existence of a minimum wavelength associated with the interatomic distances. In the case of a finite sample, there will also be a minimum frequency because there is a maximum wavelength associated with the linear size of the sample.

The minimum frequency $\omega_{\min}$ is obtained taking the wavelength along the edge of the cube as maximum half-wavelength:

$$\frac{\lambda_{\max}}{2} = (n_L - 1)a_0 = V^{1/3} \qquad (11a)$$

$$\omega_{\min} = \frac{\pi c}{V^{1/3}} \qquad (11b)$$

where $c$ is the speed of sound in the lattice, given by [5]: $c = a_0(K/M)^{1/2}$. The same speed is presumed for the transverse and longitudinal waves. It should be remarked that this is a strong assumption of our toy model made in order to keep the calculations as simple as possible.

To establish the maximum frequency, or Debye frequency $\omega_D$, the procedure is similar to the usual treatment: the integral for the density of states between $\omega_{\min}$ and $\omega_D$ has to be equal to the number of degrees of freedom for the system. Here, attention should be paid to the atoms in the surface that, consequently, have fewer degrees of freedom than those inside the cube. A careful count [3] shows that the number of oscillators (the "springs" portrayed in **Figure 1**) is: $3(n_L^3 - n_L^2) = 3(N_A - N_A^{2/3})$. Finally, having in mind the expression for the density of states $D(\omega)$ for phonons [5], we can write:

$$\int_{\omega_{\min}}^{\omega_D} D(\omega)\,\mathrm{d}\omega = \int_{\omega_{\min}}^{\omega_D} \frac{3V}{2\pi^2 c^3}\omega^2\,\mathrm{d}\omega = 3\left(N_A - N_A^{2/3}\right) \qquad (12)$$

This gives:

$$\omega_D^3 = \frac{6c^3 N_A \pi^2}{V} + \frac{c^3\pi^3}{V} - \frac{6c^3 N_A^{2/3}\pi^2}{V} \qquad (13)$$

It is interesting to analyze each term in the previous expression. The first term is the one usually present in the treatment of the Debye solid. The second is as-sociated with the existence of a minimum frequency due to the size of the sample. And the third accounts for the existence of a surface. It becomes clear that the effect caused by the finite size of the sample is negligible in the presence of the other two terms that explicitly depend on $N_A$. Now, the problem is reduced to calculating the specific heat of this nanoparticle keeping in mind the linearized dispersion relation and the existence of a minimum and a maximum frequency.

Taking into account the density of states for phonons $D(\omega)$ and expression (6), the heat capacity in the Debye approximation is:

$$C_{\text{Debye}} = \frac{9N_A\hbar^2}{\omega_D^3 k_B T^2}\int_{\omega_{\min}}^{\omega_D}\frac{\omega^4 e^{\frac{\hbar\omega}{k_B T}}}{\left(e^{\frac{\hbar\omega}{k_B T}}-1\right)^2}\,d\omega \tag{14}$$

To enable comparisons with the previous results, it is convenient to rewrite the previous expression in terms of the dimensionless temperature $t$ defined earlier, and to divide by the number of atoms. The specific heat per atom $c_{\text{Debye}}$ in the Debye approximation is obtained as follows:

$$c_{\text{Debye}} = \frac{9k_B}{u_D^3 t^2}\int_{u_{\min}}^{u_D}\frac{u^4 e^{u/t}}{\left(e^{u/t}+1\right)^2}\,du \tag{15}$$

The integration limits are $u_{\min} = \omega_{\min}/\omega_0$ y $u_D = \omega_D/\omega_0$. It is useful to write these magnitudes in terms of our model parameters, as seen in **Figure 1**. Thus, we find that:

$$u_D = \left(\frac{3n_L^3\pi^2}{4(n_L-1)} - \frac{3n_L^2\pi^2}{4(n_L-1)} + \frac{\pi^3}{8(n_L-1)}\right)^{1/3} \tag{16a}$$

and:

$$u_{\min} = \frac{\pi}{2(n_L-1)} \tag{16b}$$

It is clear that expression (15) depends only on the number of atoms $n_L$ and the reduced temperature $t$.

In **Figure 3**, the specific heat obtained with the Debye approximation is compared with that obtained exactly for a cubic nanocrystal.

The most readily visible conclusion is that the Debye approximation is unsuited for systems with tens of particles **Figure 3(a)**, as well as for systems with a few thousand atoms **Figure 3(b)** or even for a million atoms **Figure 3(c)**.

It is useful to compare the exact specific heat of a nanocrystal with the usual value of the Debye solid in the thermodynamic limit [4] [5] because, as it is known, the Debye model is a reasonable approximation for the bulk specific heat. The result is plotted in **Figure 4**. It is clear that, the specific heat of a nanocrystal is higher than the bulk value, as observed in most experiments [7] [8] [9] [10]. That is to say, our simple model explains a characteristic observed repeatedly in nanocrystals and nanoclusters.

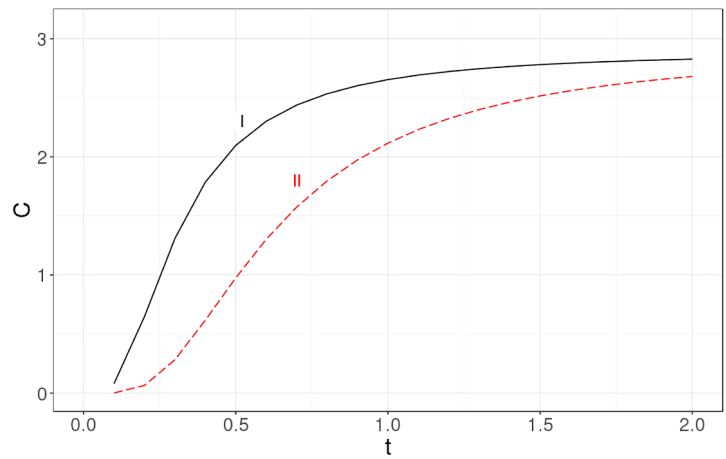

(a)

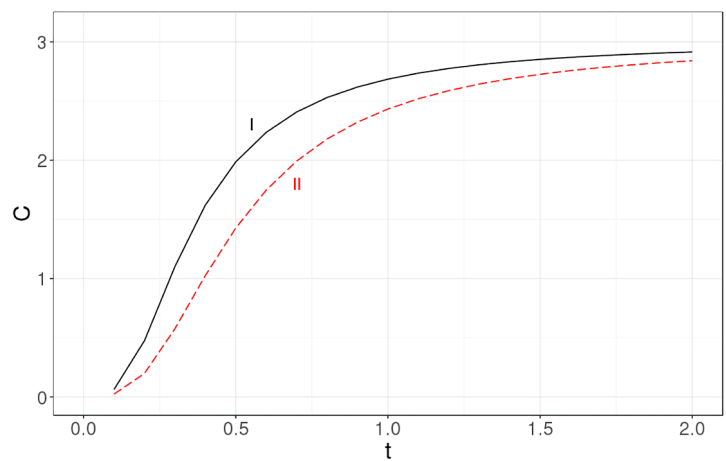

(b)

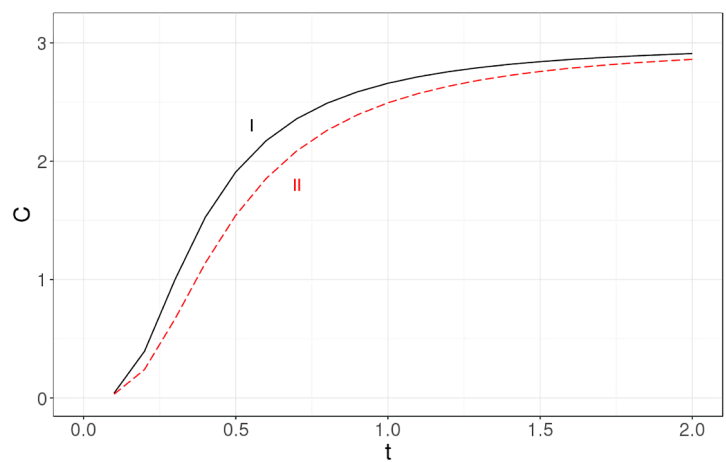

(c)

**Figure 3.** The specific heat calculated exactly (solid line labeled as I) is compared with the specific heat obtained using the Debye approximation (dashed line labeled as II) for three nanocrystal sizes: (a) $3^3$, (b) $10^3$ and (c) $100^3$. It becomes clear that the Debye approximation is unsuited for a system with tens of atoms. Neither is it a good approximation for a nanocrystal with $10^3$ or $10^6$ atoms. The specific heat is in units of $k_B$ and the temperature is a dimensionless variable as explained in the text.

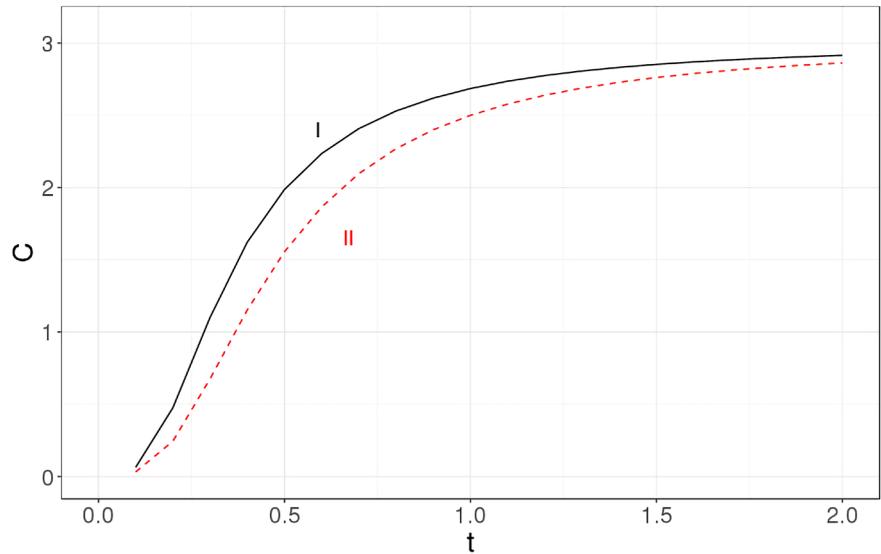

**Figure 4.** The specific heat calculated exactly (solid line labeled as I) for a nanocrystal with $10^3$ atoms is compared with the usual value predicted by the Debye model (dashed line labeled as II) in the thermodynamic limit. Clearly, our model predicts that the specific heat of a nanocrystal is higher than the bulk value, as seen in most experiments.

And how do these results compare with the Einstein solid? In this case, all the phonons have the same frequency $\omega_0$ or, more precisely, the density of states is delta function center in that frequency: $D(\omega) = \delta(\omega - \omega_0)$. Therefore, expression (6) is reduced to [3]:

$$C_{\text{Einstein}} = 3\left(N_A - N_A^{2/3}\right)\frac{\hbar^2\omega_0^2 e^{\frac{\hbar\omega_0}{k_B T}}}{k_B T^2\left(e^{\frac{\hbar\omega_0}{k_B T}} - 1\right)^2} \tag{17}$$

Naturally, it is convenient to divide by the total number of atoms $N_A = n_L^3$ and to introduce the dimensionless temperature $t$. It is necessary to keep in mind that the characteristic frequency was defined as $(K/M)^{1/2}$ in [3], but in this article it has been defined as twice that value—see Equation (4). As a result, when expressing Equation (17) in terms of the dimensionless temperature, an additional factor 2 appears. Thus, the specific heat per atom in the Einstein approximation is expressed in $k_B$ units as:

$$c_{\text{Einstein}} = 3\left(1 - \frac{1}{n_L}\right)\frac{4e^{2/t}}{t^2\left(e^{2/t} - 1\right)^2} \tag{18}$$

**Figure 5** shows $c_{ex}$, in the Debye approximation $c_{\text{Debye}}$ and in the Einstein approximation $c_{\text{Einstein}}$.

**Figure 5** is very revealing: the usual models for solids *are not applicable to nanocrystals with hundreds or thousands of atoms.*

It is reasonable to wonder why the Debye approximation is unsuited for systems with hundreds or thousands of atoms. A first conjecture is related to the finite size of the sample, which is reflected in the existence of a lower limit different from zero in the integral; however, this is not the case. The existence of a lower limit in the frequency has a negligible contribution. For example, when in the integral in (15) the lower limit is replaced by 0, the specific heat changes from 1.4325 to 1.4324 for $n_L = 10$ and $t = 1$. So, clearly, the existence of a lower integration limit different from zero is not relevant.

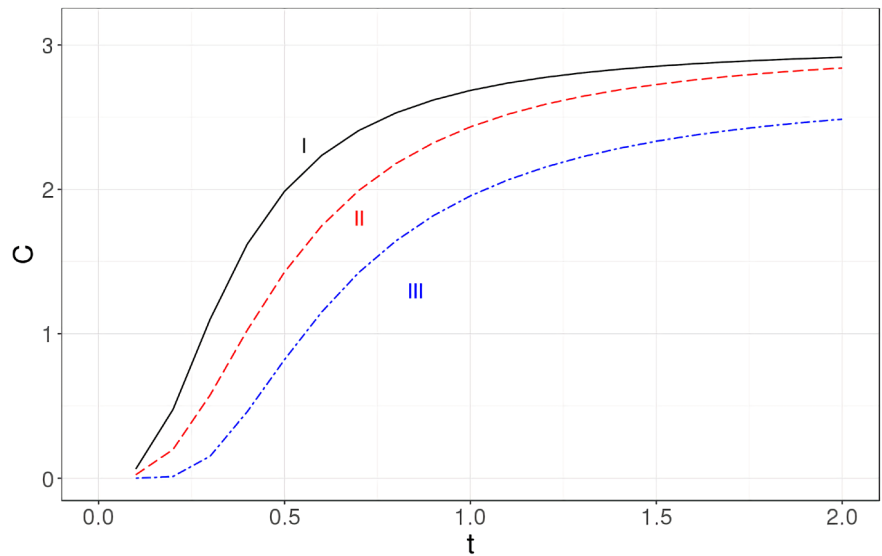

**Figure 5.** The specific heat for a cubic lattice with $10^3$ atoms is evaluated exactly (solid line labeled as I), using the Debye approximation (dashed line labeled as II) and using the Einstein model (dot-dashed line labeled as III).

Another possible cause is the linearization of the dispersion relation. Figure 6 shows the specific heat evaluated using the correct dispersion relation incorporated to expression (8), and using the linearized version as it appears in expression (10). Clearly the linearization is partially responsible for the discrepancy but other causes should be also considered.

The cause of the discrepancy seems to be twofold. On the one hand, the assumption of a continuous frequency distribution for the phonons is appropriate in the thermodynamic limit, but it is a very strong hypothesis when the number of phonons is in the order of $n_L^3$. And on the other hand, the existence of a surface seems to have a crucial effect, as already demonstrated in the case of the Einstein nanocrystal [3]. In this paper we have worked with pinned boundary conditions and this implies that surface phonons are neglected. The main difference between a macroscopic piece of material and a nanocrystal is the appearance of surface vibrational modes and a simple, straightforward application of the Debye model does not take into account such modes.

Finally, it should be noted that this model predicts that, for nanocrystals, the specific heat goes exponentially to zero, as inferred from expression (8). And this behavior is different from that observed in macroscopic crystals. There are no experimental data of the nanocrystal specific heat behavior for very low temperatures to confirm or rebut this prediction.

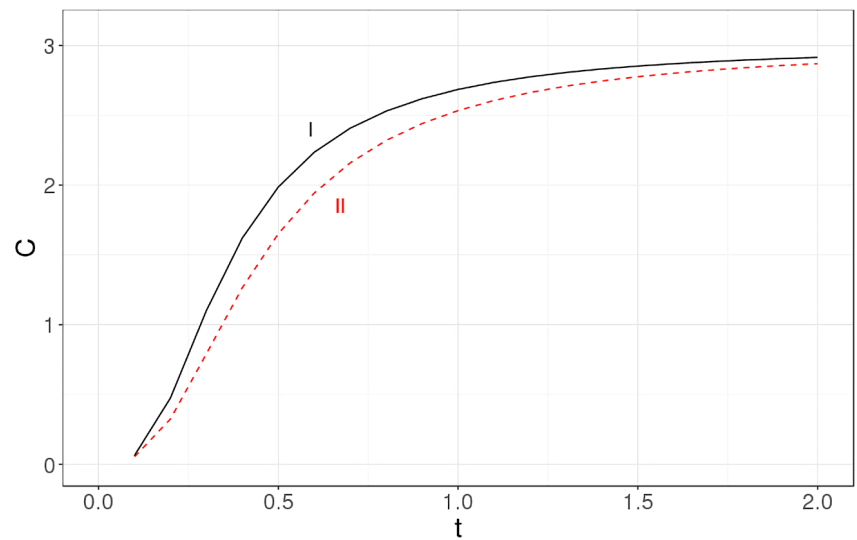

**Figure 6.** The specific heat of a nanocrystal of $10^3$ atoms evaluated using expression (8) that includes the exact dispersion relation (solid line label as I) and using the linear dispersion relation (dashed line labeled as II). The linearization of the dispersion relation seems to be partially responsible of the discrepancy between the exact value and the Debye approximation for the specific heat.

## 4. Conclusions

In this article, we have explored the validity of the Debye model at the nanoscale. For this, we have considered a very straightforward nanocrystal model: a simple cubic structure whose atoms are assumed to interact through a harmonic potential, as portrayed in **Figure 1**. This system is simple enough to allow for an exact solution, which gives the dispersion relation for the mechanical waves of the lattice—Equation (3). Thus, by thinking of the system on quantum terms, the exact dispersion relation of the phonons becomes known, and the exact specific heat can be calculated—Equation (8). The form of this specific heat coincides with that expected for a group of oscillators, though its value is higher than that found in the thermodynamic limit. **Figure 2** shows the specific heat evaluated at a fixed temperature in relation to the number of atoms in the nanocrystal; it is clear how, as the size increases, the specific heat decreases. This means that our model predicts that the specific heat is higher for nanocrystals than for crystals of macroscopic size, which is what is observed experimentally.

Once the exact specific heat is calculated for our nanocrystal, we apply the Debye approximation to allow for comparisons. To do this, the usual steps are followed, but keeping in mind that the nanocrystal has a finite size and, therefore, there is a maximum wavelength, or rather a minimum frequency. So we obtain an analytical expression for the Debye frequency—Equation (13)—where three terms are easily identified: the usual term and the corrections on account of the finite size and the surface. The first correction has proven to be irrelevant, but not the second. This agrees with our findings for the Einstein model in a previous paper [3]. The result of the current article is summarized in **Figure 5**, which shows the specific heat calculated exactly, as well as after Debye and after

Einstein. The plot makes it clear that the usual models approximate poorly the exact specific heat. Even for a nanocrystal with $10^6$ atoms, the exact value and that obtained with the Debye approximation differ considerably. The surface vibrational modes might be responsible for this disagreement. The message is then: do not forget the surface when dealing with nanocrystals. A naïve application of bulk results to a nanoparticle would lead to wrong conclusions for two reasons: 1) the surface introduces a non negligible effect (the surface phonons in the case of specific heat); 2) as the size of the particle is decreased, the approximation of using a smoothed density of states becomes less and less accurate.

The final conclusion of this article is that the standard Debye model is not very suited to model the specific heat of nanocrystals. But we also think there is a path for doing this better. Although nanocrystals with a simple cubic structure do not exist in nature, the steps outlined here could be replicated for a realistic crystal structure. It would simply be a matter of finding the frequencies of the normal modes for that realistic structure assuming that the atoms interact through a harmonic potential ("springs"). And once the formula linking the wave vector with the frequency is known, evaluating the exact specific heat becomes an exercise of statistical mechanics.

## Conflicts of Interest

The authors declare no conflicts of interest regarding the publication of this paper.